\pdfoutput=1

\documentclass[conference]{IEEEtran}
\usepackage{graphics}
\usepackage{times}
\usepackage{graphicx,color}
\usepackage{epsfig}
\usepackage[hyphens]{url}
\usepackage{hyperref}
\usepackage{flushend}
\usepackage{paralist}
\newcommand{\subparagraph}{}
\usepackage{authblk}
\usepackage{amsmath}
\usepackage{listings}
\usepackage[numbers,sort&compress]{natbib}

\usepackage{clrscode3e}
\RequirePackage{graphics}
\usepackage{xspace}
\usepackage{program}

\usepackage{marginnote}
\usepackage[small,it]{caption}

\newcommand{\paragraphbe}[1]{\vspace{0.7ex}\noindent{\bf \em #1} }

\newcommand{\ignore}[1]{}

\newcommand{\rtfifo}{{SCHED\_FIFO}\xspace}

\newcommand{\definition}[1]{{\bf #1}}

\newcommand{\perfevents}{\lstinline{perf\_events}\xspace}

\newcommand{\fixedtimebegin}{\lstinline{fixed\_time\_begin}\xspace}
\newcommand{\fixedtimeend}{\lstinline{fixed\_time\_end}\xspace}

\newcommand{\fixedtimeannotation}{\lstinline{FIXED\_TIME}\xspace}

\newcommand{\fixedtimerecord}{\lstinline{fixed\_time\_record.sh}\xspace}

\title{Robust and Efficient Elimination of Cache and Timing Side Channels}

 \author[1]{Benjamin A. Braun}
 \author[1]{Suman Jana}
 \author[1]{Dan Boneh} 
 \makeatletter
 \affil[1]{Stanford University}
 \makeatother
 \date{}
\makeatletter
\def\@copyrightspace{\relax}
\makeatother

\begin{document}

\maketitle

\begin{abstract}
Timing and cache side channels provide powerful attacks against 
many sensitive operations including cryptographic implementations. 
Existing defenses cannot protect against all classes of such 
attacks without incurring prohibitive performance overhead.
A popular strategy for defending against all classes of these attacks is to modify 
the implementation so that the timing and cache access patterns 
of every hardware instruction is independent of the secret inputs. 
However, this solution is architecture-specific, brittle, and difficult to get right.
In this paper, we propose and evaluate a robust low-overhead technique 
for mitigating 
timing and cache channels. Our solution 
requires only minimal source code changes and works across multiple languages/platforms. 
We report the experimental results of applying our solution to protect several C, C++, and Java programs.  
Our results demonstrate that our solution successfully eliminates the timing and 
cache side-channel leaks while incurring significantly lower performance overhead than 
existing approaches.
\end{abstract}

\section{Introduction}

Defending against cache and timing side channel attacks is known to be
a hard and important problem.  Timing and cache attacks can
be used to extract cryptographic secrets from running
systems~\cite{kocher1996timing,brumley07,percival05,brumley11,tromer2010efficient,percival05, gullasch2011cache,
  osvik2006cache}, spy on Web user
activity~\cite{bortz07}, and even undo the privacy of differential
privacy systems~\cite{fuzzdb,diffpriv}.  Attacks
exploiting timing side channels have been demonstrated for both remote
and local adversaries. A remote attacker is separated from its target by a
network~\cite{kocher1996timing,brumley07,percival05,brumley11}
while a local attacker can execute unprivileged spyware on the target
machine~\cite{percival05,bernstein2005cache,BM06,yarom2014recovering,barthesystem,zhang2011predictive}.

Most existing defenses against cache and timing attacks only protect against 
a subset of attacks and incur significant performance overheads. For example, one way to defend against remote timing 
attacks is to make sure that the timing of any externally observable events 
are independent of any data that should be kept secret.
Several different strategies have been proposed to achieve this, 
including application-specific changes~\cite{konighofer2008fast, blomer2005provably, kasper2009faster}, 
static transformation~\cite{van2012compiler, coppens2009practical}, 
and dynamic padding~\cite{askarov2010predictive, zhang2011predictive,
Cock_GMH_14, kopf2009provably, fuzzdb}. However, none of these strategies 
defend against local timing attacks where the
attacker spies on the target application by measuring the target's
impact on the local cache and other resources.  Similarly, the strategies for
defending against local cache attacks like static partitioning of
resources~\cite{kim2012system, ristenpart2009hey, wang2007new,
  wang2008novel}, flushing state~\cite{zhang13}, obfuscating cache
access patterns~\cite{bernstein2005cache,
  blomer2005provably,brickell2006software, tromer2010efficient,
  osvik2006cache}, and moderating access to fine-grained
timers~\cite{martin2012timewarp, li2013mitigating,vattikonda2011eliminating}, also incur significant performance 
penalties while still leaving the target potentially vulnerable to timing attacks. 
We survey these methods
in related work (Section~\ref{sec:related}).

\medskip
A popular approach for defending against both local and remote timing
attacks is to ensure that the low-level instruction sequence does not
contain instructions whose performance depends on secret
information. This can be enforced by manually re-writing the code, as was
done in OpenSSL\footnote{In the case of RSA private key operations,
  OpenSSL uses an additional defense called {\em blinding}.}, or by
changing the compiler to ensure that the generated code has this
property~\cite{coppens2009practical}.

Unfortunately, this popular strategy can fail to ensure security for several
reasons. First, the timing properties of instructions may differ in
subtle ways from one architecture to another (or even from one processor model to another) 
resulting in an instruction sequence that is unsafe for some architectures/processor models. Second,
this strategy does not work for languages like Java where the Java Virtual Machine (JVM) 
optimizes the bytecode at runtime and may inadvertently introduce
secret-dependent timing variations.  Third, manually ensuring that a
certain code transformation prevents timing attacks can be extremely difficult and tedious,
as was the case when updating OpenSSL to prevent the Lucky-thirteen
timing attack~\cite{AGLLucky13}.

\paragraphbe{Our contribution.}
We propose the first low-overhead, application-independent, and  cross-language defense 
that can protect against both local and remote timing attacks with minimal
application code changes. We show that our defense is
language-independent by applying the strategy to protect applications
written in Java and C/C++. Our defense requires relatively simple
modifications to the underlying OS and can run on off-the-shelf
hardware. 

We implement our approach in Linux and show that the execution times of protected functions 
 are independent of secret data. We also demonstrate that the performance overhead of our defense
  is low. For example, the performance overhead to protect the entire state machine running inside a 
  SSL/TLS server against all known timing- and cache-based side channel
  attacks is less than $5\%$ in connection latency.

We summarize the key insights behind our solution (described in detail in Section~\ref{solution}) below.
\begin{itemize}
\item
We leverage programmer code annotations to identify and protect sensitive code that operates on secret data. 
Our defense mechanism only protects the sensitive functions. This lets us minimize the performance impact of 
our scheme by leaving the performance of non-sensitive functions unchanged.

\item We further minimize the performance overhead by separating and accurately accounting for secret-dependent 
  and secret-independent timing variations. Secret-independent timing variations (e.g., the ones caused by interrupts, 
  the OS scheduler, or non-secret execution flow) do not leak any  sensitive information to the attacker and thus 
  are treated differently than secret-dependent variations by our scheme.

\item 
  We demonstrate that existing OS services like schedulers and hardware features like memory hierarchies 
  can be leveraged to create a lightweight isolation mechanism that can protect a sensitive function's execution 
  from other local untrusted processes and minimize timing variations during the function's  execution. 

\item
 We show that naive implementations of delay loops in most existing hardware leak timing information due 
 to the underlying delay primitive's (e.g., NOP instruction) limited accuracy. We create and evaluate a new 
 scheme for implementing delay loops that prevents such leakage while still using existing coarse-grained 
 delay primitives. 
    
\item  
  We design and evaluate a lazy state cleansing mechanism that clears the sensitive state left in shared 
  hardware resources (e.g., branch predictors, caches, etc.) before handing them over to an untrusted 
  process. We find that lazy state cleansing incurs significantly less overhead than performing state cleaning 
  as soon as a sensitive function finishes execution.

\end{itemize}

\section{Known timing attacks}

Before describing our proposed defense we briefly survey different
types of timing attackers.  In the previous section, we discussed the
difference between a local and a remote timing attacker: a local
timing attacker, in addition to monitoring the total computation time, 
can spy on the target application by monitoring the state of shared hardware 
resources such as the local cache.

\paragraphbe{Concurrent vs. non-concurrent attacks.}
In a concurrent attack, the attacker can probe shared resources while
the target application is operating.  For example, the attacker can
measure timing information or inspect the state of the shared resources at intermediate steps of a sensitive operation. The attacker's process can control the
concurrent access by adjusting its scheduling parameters and its core
affinity in the case of symmetric multiprocessing (SMP).

A non-concurrent attack is one in which the attacker only gets 
to observe the timing information or shared hardware state at the 
beginning and the end of the sensitive computation. For example, 
a non-concurrent attacker can extract secret information using only 
the {\em aggregate time} it takes the target application to process 
a request.

\paragraphbe{Local attacks.} Concurrent local attacks are the most prevalent class of timing
attacks in the research literature. Such attacks are known to be able
to extract the secret/private key against a wide-range of ciphers 
including RSA~\cite{percival05, aciicmez2007fpu},
AES~\cite{tromer2010efficient, yarom14, gullasch2011cache,
  osvik2006cache}, and ElGamal~\cite{zhang2012cross}. These attacks
exploit information leakage through a wide range of shared hardware
resources: L1 or L2 data
cache~\cite{tromer2010efficient,percival05, gullasch2011cache,
  osvik2006cache}, L3 cache
\cite{yarom14,irazoquijackpot}, instruction cache~\cite{icache, zhang2012cross},
branch predictor cache~\cite{branch_predictor1, branch_predictor2},
and floating-point multiplier~\cite{aciicmez2007fpu}.  

There are several known local non-concurrent attacks as well. 
Osvik et al.~\cite{osvik2006cache}, Tromer et
al.~\cite{tromer2010efficient}, and Bonneau and Mironov~\cite{BM06}
present two types of local, non-concurrent attacks against AES
implementations. In the first, {\it prime and probe}, the attacker
``primes'' the cache, triggers an AES encryption, and 
``probes'' the cache to learn information about the AES private
key. The spy process primes the cache by loading its own memory content 
into the cache and probes the cache by measuring the time to reload the
memory content after the AES encryption has completed.  This attack involves
the attacker's spy process measuring its own timing information to
indirectly extract information from the victim application.
Alternatively, in the {\it evict and time} strategy, the attacker
measures the time taken to perform the victim operation, evicts
certain chosen cache lines, triggers the victim operation and measure
its execution time again. By comparing these two execution times, the
attacker can find out which cache lines were accessed during the
victim operation.  Osvik et al. were able to extract an 128-bit AES
key after only 8,000 encryptions using the prime and probe attack.

\paragraphbe{Remote attacks.}  
All existing remote attacks~\cite{kocher1996timing,brumley07,percival05,brumley11} 
are non-concurrent, however this is not fundamental. A hypothetical remote, yet 
concurrent, attack would be one in which the remote attacker submits requests 
to the victim application at the same time that another non-adversarial client
sends some requests containing sensitive information to the victim application. The attacker may then be 
able to measure timing information at intermediate steps  of the non-adversarial client's 
communication with the victim application and infer the sensitive content.

\section{Threat Model}
\label{s:threat_model}

We allow the attacker to be local or remote and to execute concurrently or 
non-concurrently with the target application. We assume that the attacker can 
only run spy processes as a different
non-privileged user (i.e., no super-user privileges) than the
owner of the target application. We also assume that the spy process cannot 
bypass the standard user-based isolation provided by the operating system. We believe 
that these are very realistic assumptions because if either one of these 
assumptions fail, the spy process can steal the user's sensitive information 
without resorting to side channel attacks in most existing operating systems.  

In our model, the operating system and the underlying hardware are trusted. 
Similarly, we expect that the attacker does not have
physical access to the hardware and cannot monitor side channels such
as electromagnetic radiations, power use, or acoustic emanations.
We are only concerned with timing and cache side channels since they are 
the easiest side channels to exploit without physical access to the victim machine.

\section{Our Solution}
\label{solution}
In our solution, developers annotate the functions performing 
sensitive computation(s) that they would like to protect. For the rest 
of the paper, we refer to such functions as \definition{protected functions}.
Our solution instruments the protected functions such that our stub code is invoked 
before and after execution of each protected function. 
The stub code ensures that the protected functions, all other functions
that may be invoked as part of their execution, and all the secrets that they
operate on are safe from both local and remote timing attacks. 
Thus, our solution automatically prevents leakage of sensitive information by all functions 
(protected or unprotected) invoked during a protected function's execution.

Our solution ensures the following properties for each protected function:  
\begin{itemize}
\item
We ensure that the execution time of a protected function as observed by either a remote or 
a local attacker is independent of any secret data the function operates on. This prevents an 
attacker from learning any sensitive information by observing the execution time of a protected 
function.

\item
We clean any state left in the shared hardware resources (e.g., caches) by a protected function 
before handing the resources over to an untrusted process. As described earlier in our threat 
model (Section~\ref{s:threat_model}), we treat any process as untrusted unless it belongs to 
the same user who is performing the protected computation. We cleanse shared state only when 
necessary in a lazy manner to minimize the performance overhead.  
    
\item
We prevent other concurrent untrusted processes from accessing any intermediate state left 
in the shared hardware resources during the protected function's execution. We achieve this by 
efficiently dynamic partitioning the shared resources while incurring minimal performance overhead.  

\end{itemize}

\begin{figure}[!hbtp]
\centering
\includegraphics[width=3.5in]{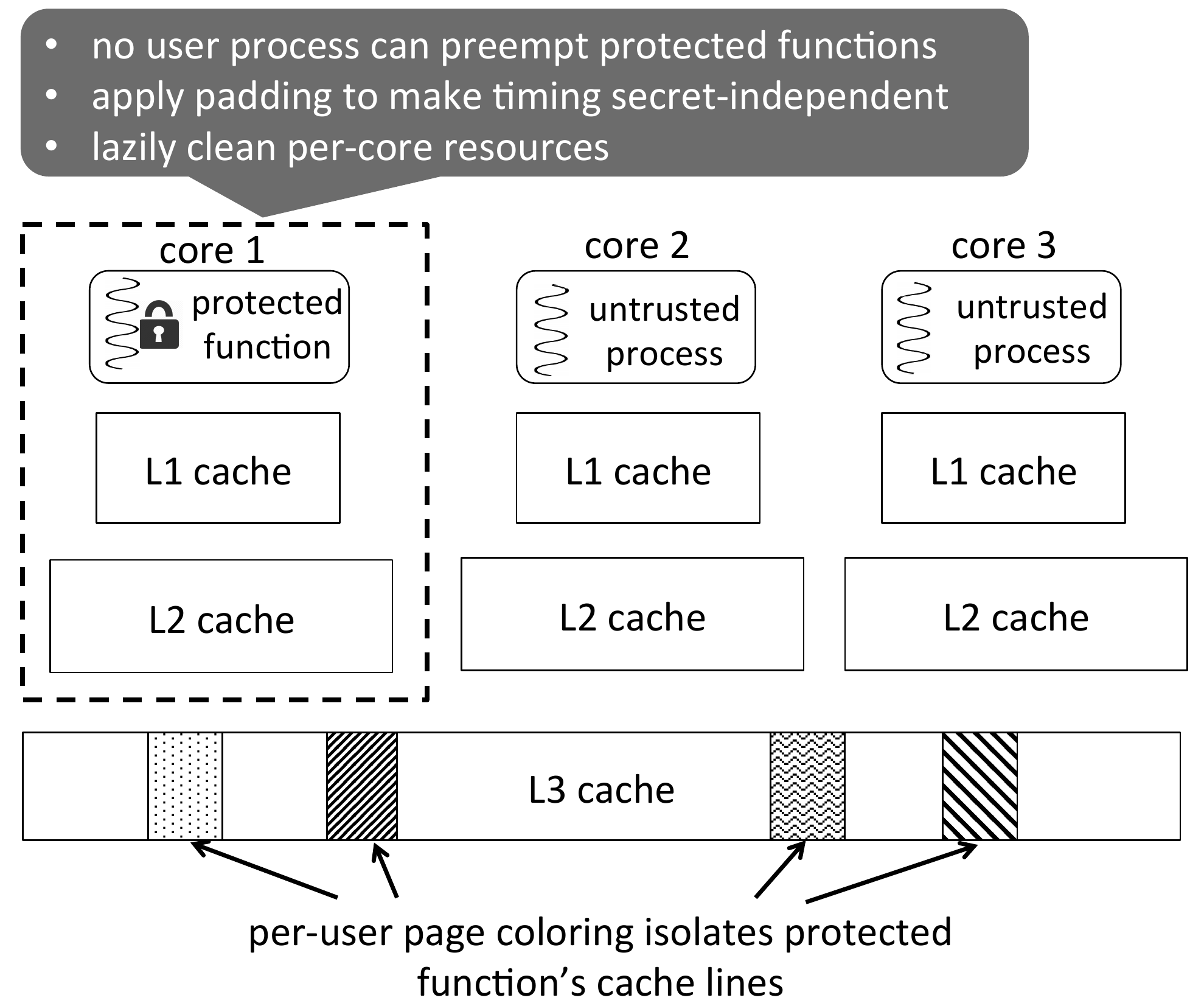}
\caption{Overview of our solution}
\label{f:fixedtime_overview}
\end{figure}

Figure~\ref{f:fixedtime_overview} shows the main components of our solution.
We use two high-level mechanisms to provide the properties described above for each protected 
function: time padding and preventing leakage through shared resources. We first briefly summarize 
these mechanisms below and then describe them in detail in Sections~\ref{time_padding}
and~\ref{sharedresource-components}.

\paragraphbe{Time padding.} 
We use time padding to make sure that a protected function's execution time does not depend 
on the secret data. The basic idea behind time padding is simple\textemdash pad the
protected function's execution time to its worst-case runtime over all possible inputs. 
The idea of padding execution time to an upper limit to prevent timing channels itself is not 
new and has been explored in several prior projects~\cite{askarov2010predictive, zhang2011predictive, Cock_GMH_14, kopf2009provably, fuzzdb}. 
However, all these solutions suffer from two major problems which prevent them from being 
adopted in real-world setting: 
\begin{inparaenum}[i)]
\item
they incur prohibitive performance overhead ($90-400$\% in macro-benchmarks~\cite{zhang2011predictive}) because they have to add
a large amount of time padding in order to prevent any timing information leakage to a 
remote attacker, and 
\item
they do not protect against {\em local} adversaries who can infer the actual unpadded 
execution time through side channels beyond network events (e.g., by 
monitoring the cache access patterns at periodic intervals).  
\end{inparaenum}

We solve both of these problems in this paper. One of our main
contributions is a new low-overhead time padding scheme that can
prevent timing information leakage of a protected function to both
local and remote attackers. We minimize the required time padding
without compromising security by adapting the worst-case time
estimates using the following three principles:

\begin{enumerate}
\item
We adapt the worst-case execution estimates to the target hardware and
the protected function. We do so by providing an offline profiling
tool to automatically estimate worst-case runtime of a particular
protected function running on a particular target hardware platform.
Prior schemes estimate the worst-case execution times for complete
services (i.e., web servers) across all possible hardware
configurations. This results in an over-estimate of the time pad that
hurts performance. 

\item  
We protect against local (and remote) attackers by ensuring that an
untrusted process cannot intervene during a protected function's
execution.  We apply time padding at the end of {\em every} protected
function's execution.  This ensures minimal overhead while preventing
a local attacker from learning the running time of protected
functions.  Prior schemes applied a large time pad before sending a
service's output over the network. Such schemes are not secure against
local attackers who can use local resources, such as cache behavior,
to infer the execution time of individual protected functions. 

\item
Timing variations result from many factors.  Some are
secret-dependent and must be prevented, while others are secret
independent and cause no harm.  For example, timing variations due to
the OS scheduler and interrupt handlers are generally harmless.  We
accurately measure and account for secret-dependent variations and 
ignore the secret-independent variations. This 
lets us compute an optimal time pad needed to protect secret data.  
None of the existing time
padding schemes distinguish between the secret-dependent and
secret-independent variations. This results in unnecessarily large
time pads, even when secret-dependent timing variations are small.
\end{enumerate}

\paragraphbe{Preventing leaks via shared resources.} 
We prevent information leakage through shared resources 
without adding significant performance overhead to
the process executing the protected function or to other
(potentially malicious) processes.
Our approach is as follows:

 \begin{itemize}
\item 
We leverage the multi-core processor architecture found in most modern processors to minimize the amount 
of shared resources during a protected function's execution without hurting performance.
We dynamically reserve exclusive access to a physical core (including all per-core caches 
such as L1 and L2) while it is executing a protected function. This ensures that a local attacker 
does not have concurrent access to any per-core resources while a protected function is accessing 
them.

\item
For L3 caches shared across multiple cores, we use page coloring to ensure that cache accesses during 
a protected function's execution are restricted within a reserved portion of the L3 cache. We further ensure 
that this reserved portion is not shared with other users' processes. This prevents the attacker from learning 
any information about protected functions through the L3 cache.

\item
We lazily cleanse the state left in both per-core resources (e.g., L1/L2 caches, branch predictors) and resources 
shared across cores (e.g., L3 cache) only before handing them over to untrusted processes. This minimizes
the overhead caused by the state cleansing operation.

\end{itemize}

\subsection{Time padding}
\label{time_padding}
We design a safe time padding scheme that defends against both local and remote attackers inferring 
sensitive information from observed timing behavior of a protected function. Our design consists of 
two main components: estimating the padding threshold and applying the padding safely without 
leaking any information. We describe these components in detail next.

\paragraphbe{Determining the padding value.}
Our time padding only accounts for secret-dependent time variations.
We discard variations due to interrupts or OS scheduler preemptions.
To do so we rely Linux's ability to keep track 
of the number of external preemptions. We adapt the total 
padding time based on the amount of time that a protected function 
is preempted by the OS.
\begin{itemize}
\item Let $T_{\text{max}}$ be the worst-case execution time of a protected 
function when no external preemptions occur.
\item Let $T_{\text{ext\_preempt}}$ be the worst-case time spent during
preemptions given the set of $n$ preemptions that occur during the 
execution of the protected function. 
\end{itemize}
Our padding mechanism pads the execution of each protected function 
to $T_{\text{padded}}$ cycles, where
\[  T_{\text{padded}} = T_{\text{ext\_preempt}} + T_{\text{max}}.   \]
This leaks the amount of preemption time to the attacker, but nothing else.
Since this is independent of the secret, the attacker learns nothing
useful.

\paragraphbe{Estimating $T_{\text{max}}$.}
Our time padding scheme requires a tight estimate of the worst-case execution time (WCET) of 
every protected function. There are several prior projects that try to estimate WCET through different 
static analysis techniques~\cite{wcet_static_1, wcet_static_2}. However, these techniques require 
precise and accurate models of the target hardware (e.g., cache, branch target buffers, etc.) which 
are often very hard to get in practice. In our implementation we use a simple dynamic 
profiling method to estimate WCET described below. Our time padding scheme 
is not tied to any particular WCET estimation method and can work with other 
estimation tools.

We estimate the WCET, $T_{\text{max}}$, through dynamic offline profiling of the 
protected function. Since this value is hardware-specific, we perform the profiling on the actual hardware
that will run protected functions. To gather profiling information, we run an application that invokes protected functions
with an input generating script provided by the application developer/system administrator. 
To reduce the possibility of overtimes occurring due to uncommon inputs, it is important that 
the script generate both common and uncommon inputs. We instrument the protected functions 
in the application so that the worst-case performance behavior is stored in a profile file. We compute 
the padding parameters based on the profiling results. 

To be conservative, we obtain all profiling measurements for the protected functions 
under high load conditions (i.e., in parallel with other application that produces significant 
loads on both memory and CPU). We compute $T_{\text{max}}$ from these measurements such 
that it is the worst-case timing bound when at most a $\kappa$ fraction of all profiling readings 
are excluded. $\kappa$ is a security parameter which provides a tradeoff between security and performance.
Higher values of $\kappa$ reduce $T_{\text{max}}$ but increase the chance of overtimes. For our prototype 
implementation we set $\kappa$ to $10^{-5}$.

\paragraphbe{Safely applying padding.}
\label{s:randomized-wait-loop}
Once the padding amount has been determined using the techniques described earlier, waiting 
for the target amount might seem easy at first glance. However, there are two major issues that 
make application of padding complicated in practice as described below. 

\begin{figure}
\centering
\includegraphics[width=.7\columnwidth]{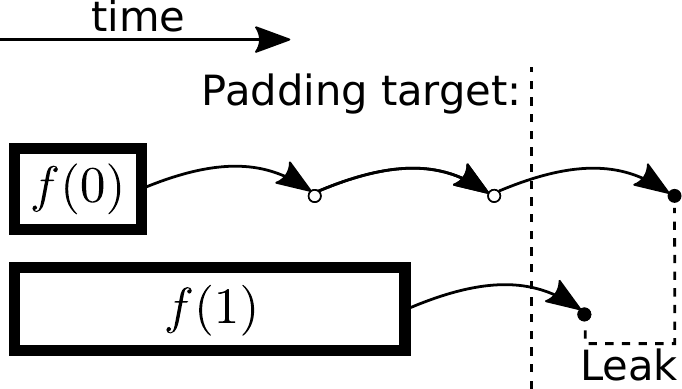}
\caption{Time leakage due to naive padding}
\label{f:naivepaddingleaks}
\end{figure}

\paragraphbe{Handling limited accuracy of padding loops.} As our solution depends on 
fine-grained padding, a naive padding scheme may leak information due to limited 
accuracy of any padding loops. Figure~\ref{f:naivepaddingleaks} 
shows that a naive padding scheme that repeatedly measures the elapsed time in a tight loop until the target
time is reached leaks timing information. This is because the loop can only break when the condition is evaluated, and hence if one iteration
of the loop takes $u$ cycles then the padding loop leaks timing information mod
$u$.  Note that earlier timing padding schemes do not get affected 
by this problem as their padding amounts are significantly larger than ours. 

Our solution guarantees that the distribution of running times of a
protected function for some set of private inputs is indistinguishable
from the same distribution produced when a different set of private inputs to the function are
used. We call this property the {\bf safe padding property}. We overcome the limitations 
of the simple wait loop by performing a timing randomization step before entering the simple wait loop. 
During this step, we perform $m$ rounds of a randomized waiting operation. This goal of this step 
is to ensure that the amount of time spent in the protected function before
the beginning of the simple wait loop, when taken modulo $u$, the stable period
of the simple timing loop (i.e. disregarding the first few iterations), is close to uniform.
This technique can be viewed as performing a random walk on the integers
modulo $u$ where the runtime distribution of the waiting operation is the support of
the walk and $m$ is the number of steps walked. Prior work by Chung et al.~\cite{chung1987random} 
has explored the sufficient conditions for the number of steps in a walk and its support that 
produce a distribution that is exponentially close to uniform.

For the purposes of this paper, we perform timing randomization 
using a randomized operation with $256$ possible inputs that runs for $X + c$
cycles on input $X$ where $c$ is a constant. We describe the details of this operation
in Section~\ref{s:implementation}. We then choose $m$ to defeat our empirical
statistical tests under pathological conditions that are very favorable
to an attacker as shown in Section~\ref{s:evaluation}.

For our scheme's guarantees to hold, the randomness used inside the randomized 
waiting operation must be generated using a cryptographically secure generator. 
Otherwise, if an attacker can predict the added random noise, she can subtract it 
from the observed padded time and hence derive the original timing signal, modulo $u$.

A padding scheme that pads to the target time $T_{\text{padded}}$ using a simple padding loop and performs 
the randomization step after the execution of the protected function will not leak any information about 
the duration of the protected function, as long as the following conditions hold: (i) no
preemptions occur; (ii) the randomization step successfully yields a distribution of runtimes that is 
uniform modulo $u$; (iii) The simple padding loop executes for enough iterations so that 
it reaches its stable period. The security of this scheme under these assumptions 
can be proved as follows.

Let us assume that the last iteration of the simple wait loop take $u$ cycles. 
Assuming the simple wait loop has iterated enough times to reach its stable period, 
we can safely assume that $u$ does not depend on when the simple wait loop started running. 
Now, due to the randomization step, we assume that the amount of time spent up to the 
start of the last iteration of the simple wait loop, taken modulo $u$, is uniformly
distributed. Hence, the loop will break at a time that is between the target time 
and the target time plus $u - 1$. Because the last iteration began when the elapsed 
execution time was uniformly distributed modulo $u$, these $u$ different cases will occur 
with equal probability. Hence, regardless of what is done within the protected function, 
the padded duration of the function will follow a uniform distribution of $u$ different 
values after the target time. Therefore, the attacker will not learn anything from 
observing the padded time of the function.

To reduce the worst-case performance cost of the randomization step, we generate 
the required randomness at the start of the protected function, before 
measuring the start time of the protected function. This means that any variability 
in the runtime of the randomness generator does not increase $T_{\text{padded}}$.

\paragraphbe{Handling preemptions occurring inside the padding loop.}
The scheme presented above assumes that no external preemptions can occur 
during the the execution of the padding loop itself. However, blocking all preemptions 
during the padding loop will degrade the responsiveness of the system. To avoid 
such issues, we allow interrupts to be processed during the execution of the 
padding loop  and update the padding time accordingly. We repeatedly update 
the padding time in response to preemptions until a ``safe exit condition'' is met 
where we can stop padding.

Our approach is to initially pad to the target value $T_{\text{padded}}$, regardless of
how many preemptions occur. We then repeatedly increase $T_{\text{ext\_preempt}}$ 
and pad to the new adjusted padding target until we execute a padding loop where 
no preemptions occur. The pseudocode of our approach is shown in 
Figure~\ref{f:pad-adapt-preempts-pseudocode}. Our technique does not 
leak any information about the actual runtime of the protected function as 
the final padding target only depends on the pattern of preemptions but 
not on the initial elapsed time before entering the padding loops.
Note that forward progress in our padding loops is guaranteed as long as 
preemptions are rate limited on the cores executing  
protected functions. 

\begin{figure}
\begin{codebox}
\zi \Comment At the return point of a protected function: 
\zi \Comment $T_{begin}$ holds the time at function start 
\zi \Comment $I_{begin}$ holds the preemption count at function start 
\li \For $j \gets 1$ \To $m$ \Do
\li     Short-Random-Delay() \End
\li $T_{target} \gets T_{begin} + T_{max}$
\li $overtime \gets 0$
\li \For $i \gets 1$ \To $\infty$ \Do
\li     $before \gets $ Current-Time$()$
\li 	\While Current-Time() $<$ $T_{target}$, re-check.
\li     \Comment Measure preemption count and adjust target
\li     $T_{ext\_preempt} \gets (\mbox{Preemptions}() - I_{begin}) \cdot T_{penalty}$
\li     $T_{next} \gets T_{begin} + T_{max} + T_{ext\_preempt} +
overtime$
\li     \Comment Overtime-detection support
\li     \If $before \ge T_{next}$ and $overtime = 0$ \Do
\li          $overtime \gets T_{overtime}$ 
\li          $T_{next} \gets T_{next} + overtime$ \End
\li     \Comment If no adjustment was made, break
\li     \If $T_{next} = T_{target}$ \Do
\li        \Return
\End
\li      $T_{target} = T_{next}$
   \End
\li \End 
\end{codebox}

\caption{Algorithm for applying time padding to a protected function's execution.}
\label{f:pad-adapt-preempts-pseudocode}

\end{figure}

The algorithm computes $T_{\text{ext\_preempt}}$ based on observed preemptions
simply by multiplying a constant $T_{\text{penalty}}$ by the number of
preemptions. Since $T_{\text{ext\_preempt}}$ should match the worst-case
execution time of the observed preemptions, $T_{\text{penalty}}$ is the
worst-case execution time of any single preemption. Like $T_{\text{max}}$,
$T_{\text{penalty}}$ is machine specific and can be determined
empirically from profiling data.

\paragraphbe{Handling overtimes.} Our WCET estimator may miss a
pathological input that causes the protected function to run for
significantly more time than on other inputs.
While we never observed this in our
experiments, if such a pathological input appeared in the wild, the
protected function may take longer than the estimated worst-case bound and
this will result in an overtime.  This leaks information: the attacker
learns that a pathological input was just processed.  We therefore
augment our technique to detect such overtimes, i.e., when the elapsed
time of the protected function, taking interrupts into
account, is greater than $T_{\text{padded}}$.

One option to limit leakage when such overtimes are detected is to
refuse to service such requests.  The system administrator can then
act by either updating the secrets (e.g., secret keys) or increasing
the parameter $T_{\text{max}}$ of the model.  

We also support updating $T_{\text{max}}$ of a protected function on
the fly without restarting the running application. The padding
parameters are stored in a file that has the same access permissions
as the application/library containing the protected function. This
file is memory-mapped when the corresponding protected function is
called for the first time. Any changes to the memory-mapped file will
immediately impact the padding parameters of all applications invoking
the protected function unless they are in the middle of applying the
estimated padding.

Note that each overtime can at most leak $log(N)$ bits of information, where $N$ is the total number of timing 
measurements observed by the attacker.  To see why, consider a string of $N$ timing observations made by 
an attacker with at most $B$ overtimes. There can be $\lt N^B$ such unique strings and thus the maximum 
information content of such a string is $\lt B log(N)$ bits, i.e., $\lt log(N)$ bits per overtime.  However, the actual effect 
of such leakage depends on how much entropy an application's timing patterns for different inputs have. For example, 
if an application's execution time for a particular secret input is significantly larger than all other inputs, even leaking 
$1$ bit of information will be enough for the attacker to infer the complete secret input.

\paragraphbe{Minimizing external preemptions.} Note that even though  $T_{\text{padded}}$ does not leak 
any sensitive information, padding to this value will incur significant performance overhead if $T_{\text{ext\_preempt}}$ 
is high due to frequent or long-running preemptions during the protected function's execution. Therefore, we minimize 
the external events that can delay the execution of a protected function. We describe the main external sources of 
delays and how we deal with them in detail below.

\begin{itemize}
\item
\paragraphbe{Preemptions by other user processes.} Under regular circumstances, execution of a
protected function may be preempted by other user processes. This can delay the execution of the 
protected function as long as the process is preempted. Therefore, we need to minimize such preemptions 
while still keeping the system usable.

In our solution, we prevent preemptions by other user processes during the execution of a protected
function by using a scheduling policy that prevents migrating the process to a different core and prevents
other user processes from being scheduled on the same core during the duration of the protected function's 
execution.

\item
\paragraphbe{Preemptions by interrupts.} Another common source of preemption is the hardware
interrupts served by the core executing a protected function. One way to solve this problem is to
block or rate limit the number of interrupts that can be served by a core while executing a protected
function. However, such a technique may make the system non-responsive under heavy load. For this
reason, in our current prototype solution, we do not apply such techniques.

Note that some of these interrupts (e.g., network interrupts) can be triggered by the attacker
and thus can be used by the attacker to slow down the protected function's execution. However, in
our solution, such an attack increases $T_{\text{ext\_preempt}}$, and hence degrades performance, but does 
not cause information leakage.

\item
\paragraphbe{Paging.} An attacker can potentially arbitrarily slow down the protected function by causing memory paging 
events during the execution of a protected function. To avoid such cases, our solution forces each process executing a protected 
function to lock all of its pages in memory and disables page swapping. As a consequence, our solution currently does not allow 
processes that allocate more memory than is physically available in the target system to use protected functions.

\item
\paragraphbe{Hyperthreading.} Hyperthreading is a technique supported by modern processor cores where
one physical core supports multiple logical cores. The operating system can independently schedule tasks on these 
logical cores and the hardware transparently takes care of sharing the underlying physical core. We observed that 
protected functions executing on a core with hyperthreading enabled can encounter large amounts of slowdown. 
This slowdown is caused because the other concurrent processes executing on the same physical core can
interfere with access to some of the CPU resources.

One potential way of avoiding this slowdown is to configure the OS scheduler to prevent any untrusted process
from running concurrently on a physical core with a process in the middle of a protected function.
However, such a mechanism may result in high overheads due to the cost of actively unscheduling/migrating a 
process running on a virtual core. For our current prototype implementation, we simply disable hyperthreading as 
part of system configuration.
\item
\paragraphbe{CPU frequency scaling.} Modern CPUs include mechanisms to change the operating frequency of
each core dynamically at runtime depending on the current workload to save power. If a core's frequency
decreases in the middle of the execution of a protected function or it enters the halt state to save power, 
it will take longer in real-time, increasing $T_{\text{max}}$. To reduce
such variations, we disable CPU frequency scaling and low-power CPU states when a core executes a protected
function.
\end{itemize}

\subsection{Preventing leakage through shared resources}
\label{sharedresource-components}

We prevent information leakage from protected functions through shared resources in two ways: isolating shared resources from other concurrent processes and 
lazily cleansing state left in shared resources before handing them over to other 
untrusted processes. Isolating shared resources of protected functions from other concurrent processes 
help in preventing local timing and cache attacks as well as improving performance by minimizing 
variations in the runtime of protected functions. 

\paragraphbe{Isolating per-core resources.} As described earlier in Section~\ref{time_padding}, we disable hyperthreading 
on a core during a protected function's execution to improve performance. This also ensures that 
an attacker cannot run spy code that snoops on per-core state while a protected function is executing. 
We also prevent preemptions from other user processes during the execution of protected function 
and thus ensure that the core and its L1/L2 caches are dedicated for the protected function.

\paragraphbe{Preventing leakage through performance counters.} 
Modern hardware often contain performance counters that keep track of different performance
events such as the number of cache evictions or branch mispredictions
occurring on a particular core. A local attacker with access to these
performance counters may infer the secrets used during a protected 
function's execution. Our solution, therefore, restricts
access to performance monitoring counters so that a user's process
cannot see detailed performance metrics of another user's processes.
We do not restrict, however, a user from using hardware performance
counters to measure the performance of their own processes.

\paragraphbe{Preventing leakage through L3 cache.}
As L3 cache is a shared resources across multiple cores, we use page coloring to 
dynamically isolate the protected function's data in the L3 cache. To support page 
coloring we modify the OS kernel's physical page allocators so that they do not allocate 
pages having any of $C$ reserved {\it secure page colors}, unless the caller specifically 
requests a secure color. Pages are colored based on which L3 cache sets a page maps 
to. Therefore, two pages with different colors are guaranteed never to conflict 
in the L3 cache in any of their cache lines.

In order to support page coloring, we disable transparent huge pages and set up access 
control to huge pages. An attacker that has access to a huge page can evade the isolation 
provided by page coloring, since a huge page can span multiple page colors. Hence, we prevent 
access to huge pages (transparently or by request) for non-privileged users.

As part of our implementation of page coloring, we also disable memory deduplication
features, such as kernel same-page merging. This prevents a
secure-colored page mapped into one process from being transparently
mapped as shared into another process. Disabling memory deduplication is not unique 
to our solution and has been used in the past in hypervisors to prevent leakage of information 
across different virtual machines~\cite{suzaki2011dedup}.

When a process calls a protected function for the first time, we invoke a kernel module routine to remap all pages allocated 
by the process in private mappings (i.e., the heap, stack, text-segment, library code, and library
data pages) to pages that are not shared with any other user's processes. We also ensure  
these pages have a page color reserved by the user executing the protected function. The remapping transparently changes
the physical pages that a process accesses without modifying the virtual
memory addresses, and hence requires no special application support.
If the user has not yet reserved 
any page colors or there are no more remaining pages of any of her reserved 
page colors, the OS allocates one of the reserved colors for the user. Also, the 
process is flagged with a "secure-color" bit. We modify the OS so that it recognizes 
this flag and ensures that the future pages allocated to a private mapping for the process 
will come from a reserved page color for the user. Note that since we only remap private mappings, we do not protect 
applications that access a shared mapping from inside a protected function. 

This strategy for allocating page colors to users has a minor potential downside that
such a system restricts the numbers of different users' processes that can concurrently 
call protected functions. We believe that such a restriction is a reasonable 
trade-off between security and performance.

\paragraphbe{Lazy state cleansing.}
To ensure that an attacker does not see the tainted state in a per-core
resource after a protected function finishes execution, we lazily delete 
all per core resources. When a protected function returns, we mark the CPU 
as ``tainted'' with the user ID of the caller process. The next time the OS attempts 
to schedule a process from a different user on the core, it will first flush all per-CPU 
caches, including the L1 instruction cache, L1 data cache, L2 cache, Branch Translation Buffer (BTB), 
and Translation lookaside buffer (TLB).  Such a scheme ensures that the 
overhead of flushing these caches can be amortized over multiple invocations 
of protected functions by the same user.

\section{Implementation}
\label{s:implementation}
We built a prototype implementation of our protection mechanism
for a system running Linux OS. We describe the different components of our implementation below. 

\subsection{Programming API}
\label{programming-api}
We implement a new function annotation \fixedtimeannotation for the C/C++
language that indicates that a function should be protected. The annotation can
be specified either in the declaration of the function or at its definition. Adding 
this annotation is the only change to a C/C++ code base that a programmer has to
make in order to use our solution. We wrote a plugin for the Clang C/C++ compiler that 
handles this annotation. The plugin automatically inserts a call to the function \fixedtimebegin at the 
start of the protected function and a call to \fixedtimeend at any return point of
the function. These functions protect the annotated function using the
mechanisms described in Section~\ref{solution}.

Alternatively, a programmer can also call these
functions explicitly. This is useful for protecting ranges of code within function
such as the state transitions of the TLS state machine (see Section~\ref{s:protect-tls-state}).
We provide a Java native interface wrapper to both \fixedtimebegin and
\fixedtimeend functions, for supporting protected functions written in Java.

\subsection{Time padding}
For implementing time padding loops, we read from the timestamp counter in x86 processors 
to collect time measurements. In most modern x86 processors, including the one we tested on,  
the timestamp counter has a constant frequency regardless of 
the power saving state of a processor. We generate pseudorandom bytes for the randomized 
padding step using the ChaCha/8 stream cipher \cite{djbchacha}.
We use a value of 300~$\mu$s for $T_{penalty}$ as this bounds the
worst-case slowdown due to a single interrupt we observed in our
experiments.

Our implementation of the randomized wait operation takes an input $X$ and simply
performs $X + c$ noops in a loop, where $c$ is a large enough value so that
the loop takes one cycle longer for each additional iteration. We
observe that $c = 46$ is sufficient to achieve this property.

Some of the OS modifications specified in our solution are implemented
as a loadable kernel module. This module supports an IOCTL call to mark a core as tainted
at the end of a protected function's execution. The module also supports an
IOCTL call that enables fast access to the interrupt and context-switch count.
In the standard Linux kernel, the interrupt count is usually accessed through the 
proc file system interface. However, such 
an interface is too slow for our purposes. Instead, our kernel module allocates a 
page of counters that is mapped into the virtual address space of 
the calling process. The task struct of the calling process also contains a pointer 
to these counters. We modify the kernel to check on every interrupt and
context switch if the current task has such a page, and if so, to increment the 
corresponding counter in that page.

\paragraphbe{Offline profiling.}
We provide a profiling wrapper script, \fixedtimerecord, that computes
worst-case execution time parameters of each protected function
as well as the worst-case slowdown on that function due to
preemptions by different interrupts or kernel tasks.

The profiling script automatically generates profiling information for all 
protected functions in an executable by running the application on different inputs.  
During the profiling process, we run a variety of 
applications in parallel to create a stress-testing environment that triggers 
worst-case performance of the protected function. 
To allow the stress testers to maximally slow down the user application,
we reset the scheduling parameters and CPU affinity of a thread at the
start and end of every protected function.
One stress tester generates interrupts at a high frequency using a simple program that generates a flood of 
UDP packets to the loopback network interface. We also run the mprime\footnote{http://www.mersenne.org/},
systester\footnote{http://systester.sourceforge.net}, and the
LINPACK benchmark\footnote{https://software.intel.com/en-us/articles/intel-math-kernel-library-linpack-download/}
to cause high CPU load and large amounts of memory contention.
\subsection{Prevent leakage through shared resources} 
\paragraphbe{Isolating a processor core and core-specific caches.} We disable hyperthreading 
in Linux by selectively disabling virtual cores. This prevents any other processes from interfering 
with the execution of a protected function. As part of our prototype, we also implement a simple 
version of the page coloring scheme described in Section~\ref{solution}. 

We prevent a user from observing hardware performance counters showing the performance
behavior of other users' processes. The \perfevents framework on Linux mediates access 
to hardware performance counters. We configure the framework to allow accessing per-CPU 
performance counters only by the privileged users. Note that an unprivileged user can still access
per-process performance counters that measure the performance of their own processes.

For ensuring that a processor core executing a protected function is not preempted by other user processes, as specified in Section~\ref{solution}, we depend on a scheduling mode that prevents other userspace processes from preempting a protected function. 
For this purpose, we use the Linux \rtfifo scheduling mode at maximum priority. 
In order to be able to do this, we allow unprivileged users to use \rtfifo at priority
99 by changing the limits in the \texttt{/etc/security/limits.conf} file. 

One side effect of this technique is that if a protected function manually yields to the 
scheduler or perform blocking operations, the process invoking the protected function 
may be scheduled off. Therefore, we do not allow any blocking operations or system 
calls inside the protected function. As mentioned earlier, we also disable paging for the 
processes executing protected functions by using the \texttt{mlockall()} system call 
with the \texttt{MCL\_FUTURE}.

We detect whether a protected function has violated the conditions of 
isolated execution by determining whether any voluntary context switches 
occurred during the protected function's execution. This usually indicates 
that either the protected function yield the CPU manually or performed 
some blocking operations.

\paragraphbe{Flushing shared resources.}
We modify the Linux scheduler to check the taint of a core before scheduling a 
user process on a processor core and to flush per-core resources if needed as described 
in Section~\ref{solution}.

To flush the L1 and L2 caches, we iteratively read over a segment of memory that is larger 
than the corresponding cache sizes. We found this to be significantly more efficient than using 
the \texttt{WBINVD} instruction, which we observed cost as much as 300 microseconds in 
our tests. We flush the L1 instruction cache by executing a large number of NOP instructions.

Current implementations of Linux flush the TLB during each context switch. Therefore, we do 
not need to separately flush them. However, if Linux starts leveraging the PCID feature of x86 
processors in the future, the TLB would have to be flushed explicitly. For flushing the BTB, we 
leveraged a ``branch slide'' consisting of alternating conditional branch and NOP instructions.

\section{Evaluation}
\label{s:evaluation}
To show that our approach can be applied to protect a wide variety of software, we 
have evaluated our solution in three different settings and found that our solution
successfully prevents local and remote timing attacks in all of these settings.
We describe the settings in detail below.

\paragraphbe{Encryption algorithms implemented in high level interpreted languages like Java.} 
Traditionally, cryptographic algorithms implemented in interpreted languages like Java have been harder to protect from timing attacks than those implemented in low level languages like C. Most interpreted languages are compiled down to machine code on-the-fly by a VM using Just-in-Time (JIT) code compilation techniques. The JIT compiler often optimizes the code non-deterministically to improve performance. This makes it extremely hard for a programmer to reason about the transformations that are required to make a sensitive function's timing behavior secret-independent. While developers writing low level code can use features such as in-line assembly to carefully control the machine code of their implementation, such low level control is simply not possible in a higher level language.

We show that our techniques can take care of these issues. We demonstrate that our defense can make the computation time of Java implementations of cryptographic algorithms independent of the secret key with minimal performance overhead.

\paragraphbe{Cryptographic operations and SSL/TLS state machine.}
Implementations of cryptographic primitives other than the public/private key 
encryption or decryption routines may also suffer from side channel attacks. 
For example, a cryptographic hash algorithm like SHA-1 takes different 
amount of time depending on the length of the input data. In fact, such timing 
variations have been used as part of several existing attacks against SSL/TLS protocols (e.g.,  Lucky 13).  Also, the time taken to perform the computation for implementing different stages of the SSL/TLS state machine may also be dependent on the 
secret key. 

We find that our protection mechanism can protect cryptographic primitives like hash functions as well as individual stages of the SSL/TLS state machine from timing attacks while incurring minimal overhead.

\paragraphbe{Sensitive data structures.}  Besides cryptographic algorithms, timing channels also occur in the context of different data structure operations like  hash table  lookups. For example, hash table lookups may take different amount of time depending on how many items are present in the bucket where the desired item is located. It will take longer time to find items in buckets with higher number of items than in the ones with less items. This signal can be exploited by an attacker to cause denial of service attacks
\cite{crosby03}.
We demonstrate that our technique can prevent timing leaks using the associative arrays in C++ STL, a popular hash table implementation.

\paragraphbe{Experiment setup.}
We perform all our experiments on a machine with 2.3GHz Intel Xeon E5-2630 
CPUs organized in $2$ sockets each containing $6$ physical cores unless otherwise specified. 
Each core has a 32KB L1 instruction cache, a 32KB L1 data cache, and a 256KB L2 cache. 
Each socket has a 15MB L3 cache. The machine has a total of 64GB of RAM.

For our experiments, we use OpenSSL version 1.0.1l and Java version BouncyCastle 1.52 (beta). The test machine runs Linux kernel version 3.13.11.4 with our modifications as discussed in Section~\ref{s:implementation}.

\subsection{Security evaluation.}
\label{sec_eval}
\paragraphbe{Preventing a simple timing attack.}
To determine the effectiveness of our safe padding technique, we
first test whether our technique can protect against a large timing channel 
that can distinguish between two different inputs of a simple function.
To make the attacker's job easier,  we craft a simple function that has an 
easily observable timing channel\textemdash the function executes a loop 
for $1$ iteration if the input is $0$ and $11$ iterations otherwise. 
We use the x86 \texttt{loop} instruction
to implement the loop and  just a single \texttt{nop} instruction as the body 
of the loop. We assume that the attacker calls the protected function directly 
and measures the value of the timestamp counter immediately before and 
after the call. The goal of the attacker is to distinguish between two different 
inputs ($0$ and $1$) by monitoring the execution time of the function. 
Note that these conditions are extremely favorable for an attacker.

\begin{figure}
\centering
\parbox{\textwidth}{
\hspace{2.35in}
\includegraphics[width=.8in,trim=3.4in 1in .3in .7in,clip=true]{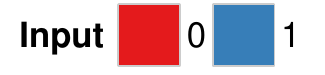}
}
\parbox{\textwidth}{\vspace{-.27in}A. Unprotected}
\parbox{\textwidth}{\vspace{-.12in}
\includegraphics[width=\columnwidth,trim=0 0 0 0,clip=true]{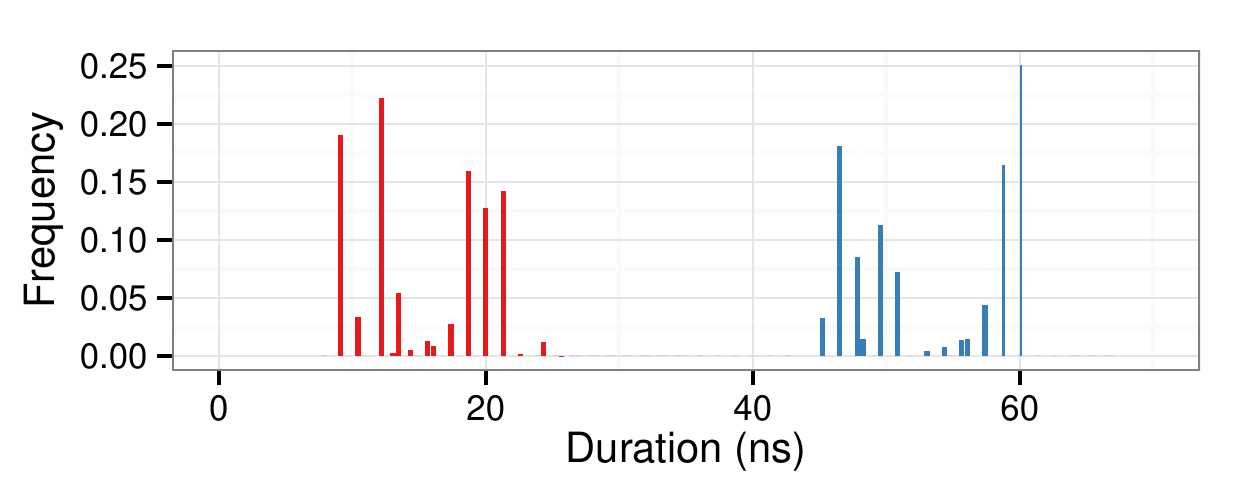}
}
\parbox{\textwidth}{B. With time padding but no randomized noise}
\includegraphics[width=\columnwidth,trim=0 0 0 0,clip=true]{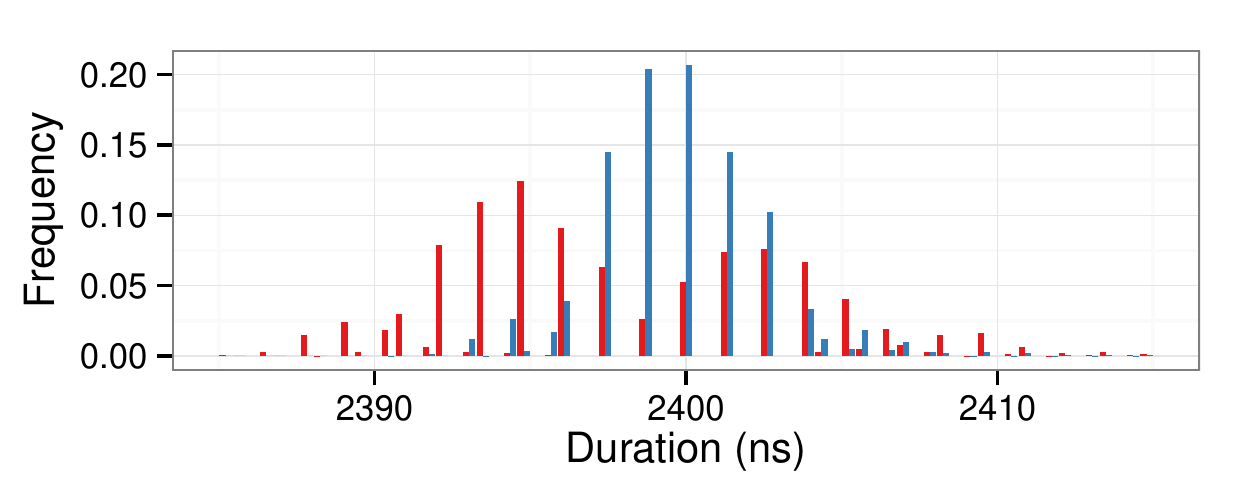}
\parbox{\textwidth}{C. Full protection (padding+randomized noise)}
\includegraphics[width=\columnwidth]{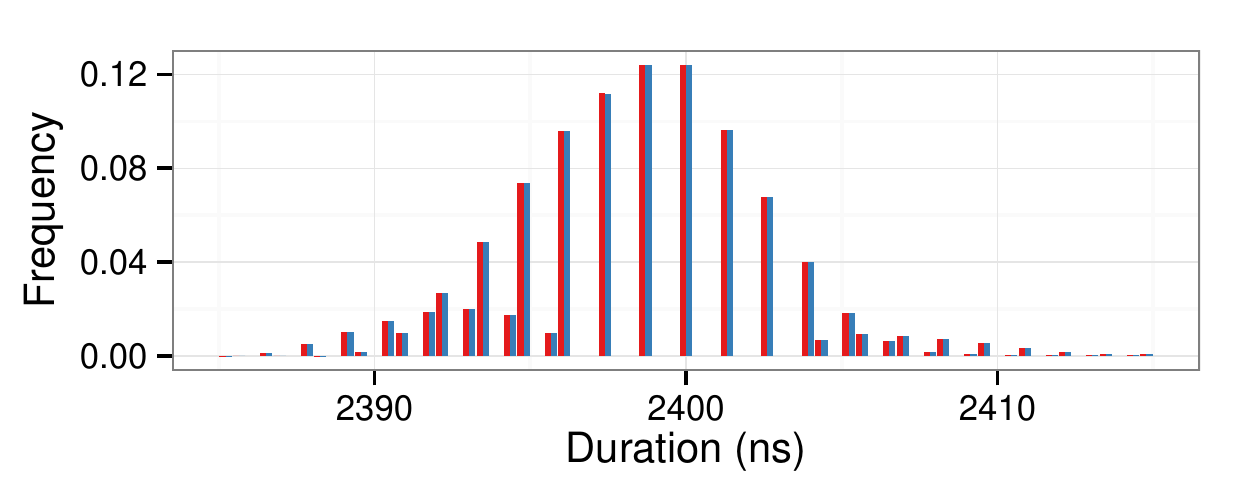}
\caption{Defeated distinguishing attack}
\label{f:distinguishing_attack_unprotected}
\end{figure}

We found that our defense completely defeats such a distinguishing attack despite the highly favorable conditions for the attacker. We also found that the timing randomization step (described in Section~\ref{time_padding}) is critical for such protection and a naive padding loop with any timing randomization step indeed leaks information.  Figure~\ref{f:distinguishing_attack_unprotected}(A)
shows the distributions of observed runtimes of the protected function on
inputs $0$ and $1$ with no defense applied. Figure~\ref{f:distinguishing_attack_unprotected}(B) 
shows the runtime distributions where padding is added to reach
$T_{max} = 5000$ cycles ($\approx 2.17~\mu s$) without
the time randomization step.
In both cases, it can be seen that the observed timing
distributions for the two different inputs are clearly distinguishable.
Figure~\ref{f:distinguishing_attack_unprotected}(C) shows the same distributions 
when $m=5$ rounds of timing randomization are applied along with time padding. 
In this case, we are no longer able to distinguish the timing distributions.

\begin{figure}[!htbp]
\centering
\includegraphics[height=2.5in]{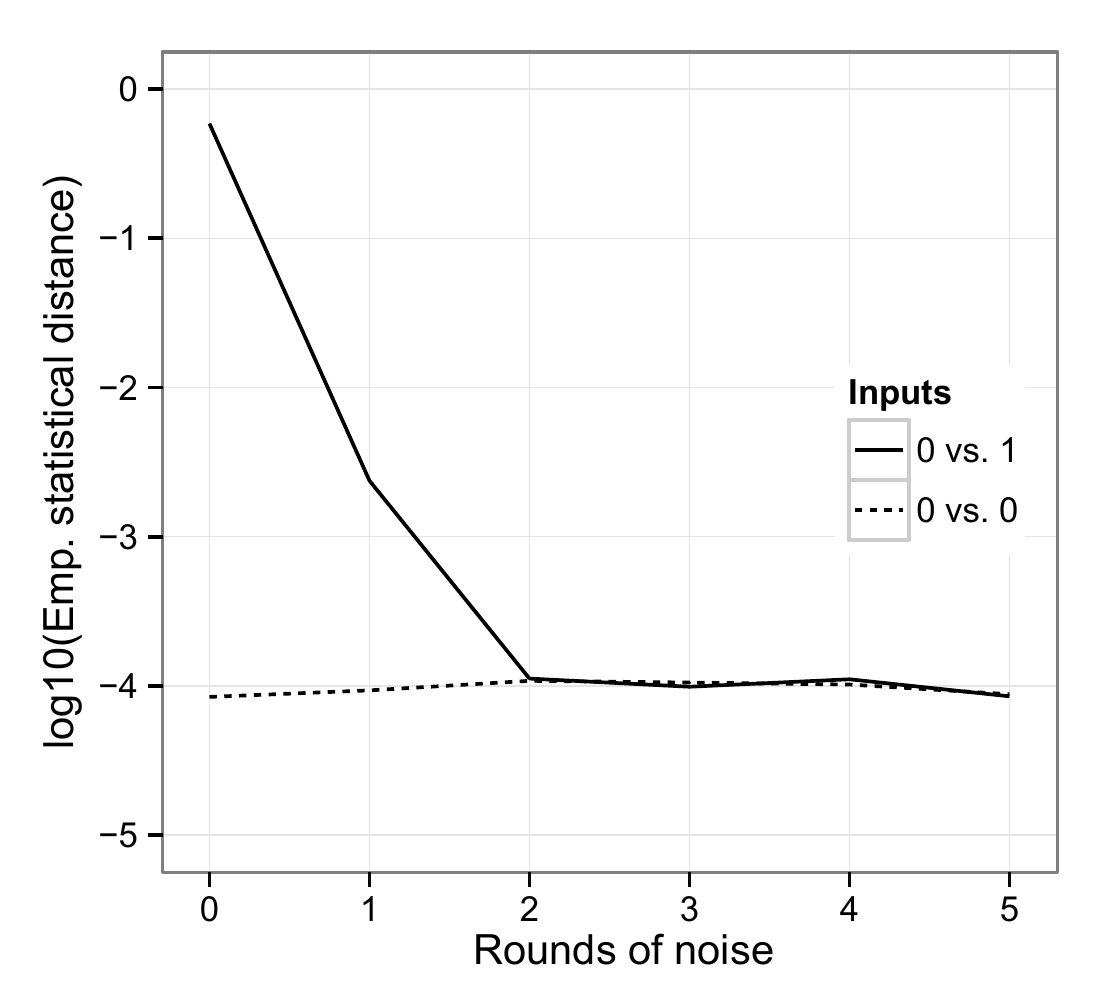}
\caption{The effect of multiple rounds of randomized noise addition on the timing channel}
\label{f:distinguishing_success}
\end{figure}

We quantify the possibility of success for a distinguishing attack in
Figure~\ref{f:distinguishing_success} by plotting the variation of empirical statistical distance 
between the observed distributions as the amount of padding noise added is changed. 
The statistical distance is computed using the following formula.
\[ d(X,Y) = \dfrac{1}{2}\sum_{i\in \Omega} \lvert P[X = i] - P[Y = i]
\rvert \]
We measure the statistical distance over the set of observations that are within 
the range of $50$ cycles on either side of the median (this contains nearly all 
observations.) Each distribution consist of around $600$ million observations.

The dashed line in Figure~\ref{f:distinguishing_success} shows the statistical 
distance between two different instances of the test function with $0$ as input. 
The solid line shows the statistical distance where one instance has $0$ as input and 
the other has $1$. We observe that the attack can be completely prevented 
if at least $2$ rounds of noise are used.

\begin{figure}
\centering
\parbox{\textwidth}{
\hspace{1.75in}
\includegraphics[height=.25in, trim=2.8in .9in .2in .75in, clip=true]{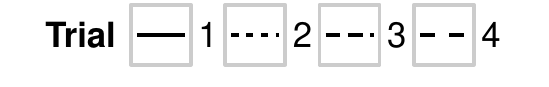}}
\parbox{\textwidth}{\vspace{-.3in}A. Unprotected}
\parbox{\textwidth}{\vspace{-.10in}
\includegraphics[height=1.3in]{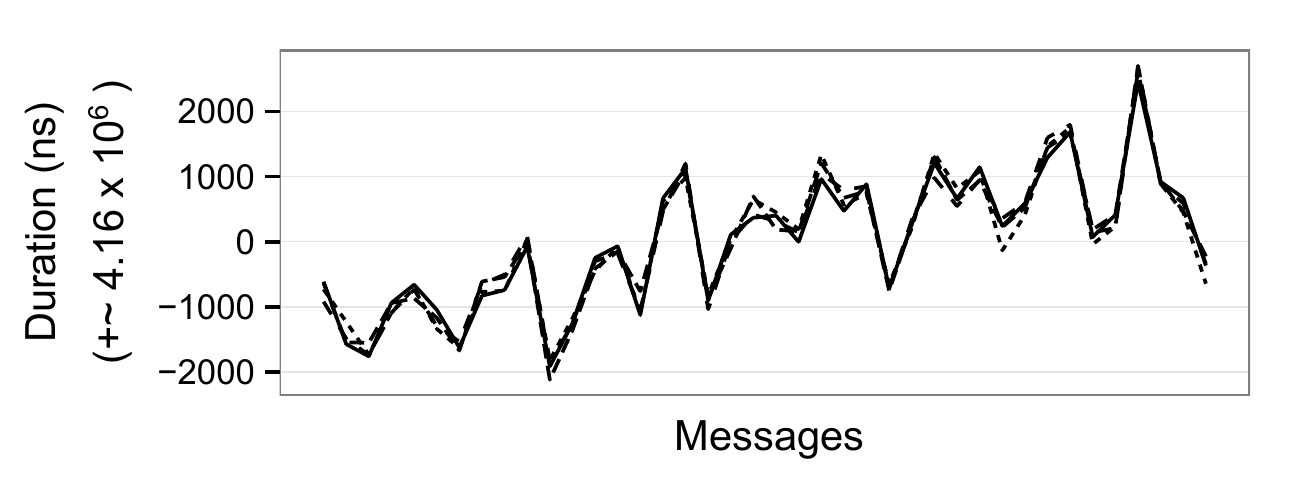}
}
\parbox{\textwidth}{B. Protected}
\includegraphics[height=1.3in]{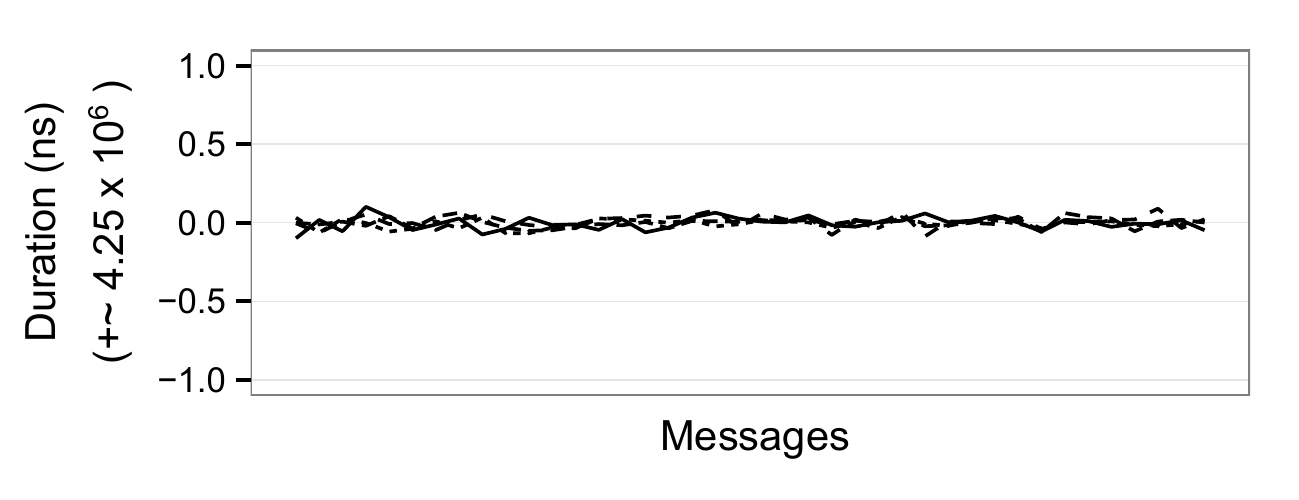}
\caption{Protecting against timing attacks on unblinded RSA}
\label{f:rsa_attack}
\end{figure}

\paragraphbe{Preventing timing attack on RSA decryption}
We next evaluate the effectiveness of our time padding approach to defeat 
the timing attack by Brumley et al.~\cite{brumley07} against unblinded RSA implementations.
Blinding is an algorithmic modification to RSA that uses randomness
to prevent timing attacks. To isolate the impact of our specific defense, we apply 
our defense to the RSA implementation in OpenSSL 1.0.1h with
such constant time defenses disabled. In order to do so, we configure OpenSSL to 
disable blinding, use the non-constant time exponentiation implementation, and use 
the non-word-based Montgomery reduction implementation.
We measure the time of decrypting $256$-byte messages with a random $2048$-bit key.
We chose messages to have Montgomery representations differing by
multiples of $2^{1016}$.
Figure~\ref{f:rsa_attack}(A) shows the average observed
running time for such a decryption operation,
which is around $4.16$ ms. The messages are displayed
from left to right in sorted order of how many
Montgomery reductions occur during the decryption. Each message was
sampled roughly $8,000$ times and the samples were randomly split into 4
sample sets. As observed by Brumley et al.~\cite{brumley07}, 
the number of Montgomery reductions can be roughly
determined from the running time of an unprotected RSA decryption.
Such information can be used to derive full length keys.

We then apply our defense to this decryption with $T_{max}$ set to
$9.68 \times 10^6$ cycles $\approx 4.21$ ms.
One timer interrupt is guaranteed to occur during such an operation, as
timer interrupts occur at a rate of $250$/s on our target machine.
We collect $30$ million measurements and observe a multi-modal padded distribution with
four narrow, disjoint peaks corresponding
to the padding algorithm using different $T_{ext\_preempt}$ values for 1, 2, 3, and 4 
interrupts respectively. The four peaks represent, respectively, $94.0\%, 5.8\%,
0.6\%,$ and $0.4\%$ of the samples.
We did not observe that these probabilities vary across different
messages.
Hence, in Figure~\ref{f:rsa_attack}(B), we show the average observed time
considering only observations from within the first peak. Again, samples
are split into $4$ random sample sets, each key is sampled around 700,000
times. We observe no message-dependent signal.

\paragraphbe{Preventing cache attacks on AES encryption.} We next verify that our
system protects against local cache attacks. Specifically, we measured 
the effectiveness of our defense against the PRIME+PROBE attack by 
Osvik et.al~\cite{osvik2006cache} on the software implementation 
of AES encryption in OpenSSL. For our tests, we apply the attack on 
only the first round of AES instead of the full AES to make the conditions very 
favorable to the attacker as subsequent rounds of AES add more noise to 
the cache readings. In this attack, the attacker first primes the cache by filling
a selection of cache sets with the attacker's memory lines. Next, the
attacker coerces the victim process to perform an AES encryption on a chosen
plaintext on the same processor core. Finally, the attacker reloads the memory lines it used to fill the
cache sets prior to the encryption. This allows the attacker to detect whether the reloaded lines 
were still cached by monitoring timing or performance counters and thus infer which memory 
lines were accessed during the AES encryption operation. 

On our test machine, the OpenSSL software AES implementation performs table lookups during the 
first round of encryption that access one of 16 cache sets in each of 4 lookup tables. The actual 
cache sets accessed during the operation are determined by XORs of the top 4 bits of
certain plaintext bytes $p_i$ and certain key bytes $k_i$. By repeatedly observing
cache accesses on chosen plaintexts where $p_i$ takes all possible values of its top 
4 bits, but where the rest of the plaintext is randomized, the attacker observes cache 
line access patterns revealing the top 4 bits of $p_i \oplus k_i$, and hence the top 4
bits of the key $k_i$. This simple attack can be extended to learn the entire
AES key.

\begin{figure}
\centering
\parbox{\textwidth}{A. Unprotected}
\parbox{\textwidth}{
\includegraphics[width=\columnwidth,trim=0 0 0 .1,clip=true]{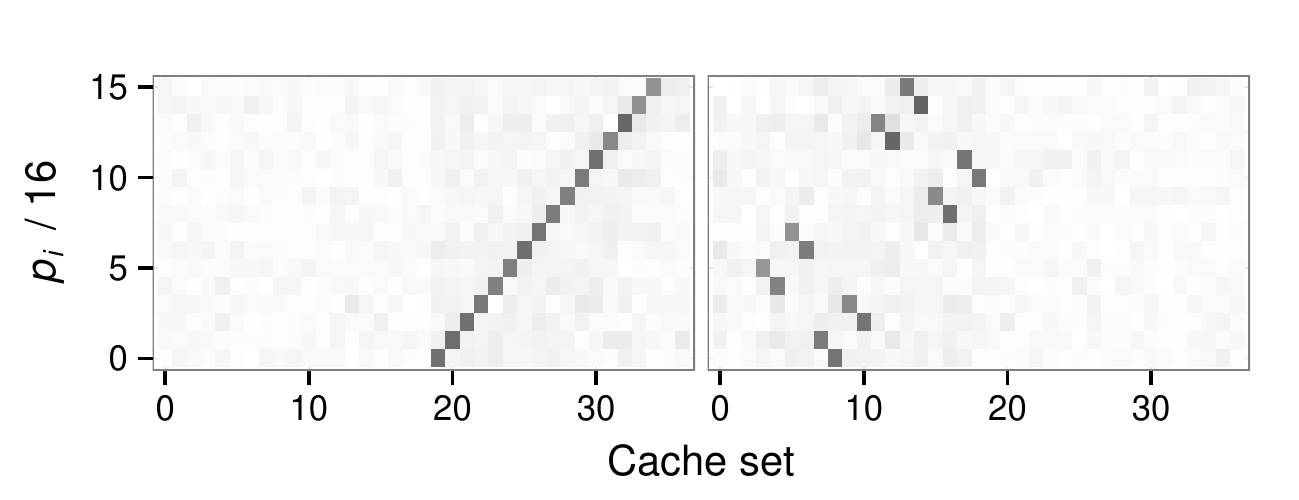}
}
\parbox{\textwidth}{B. Protected}
\includegraphics[width=\columnwidth,trim=0 0 0 .1,clip=true]{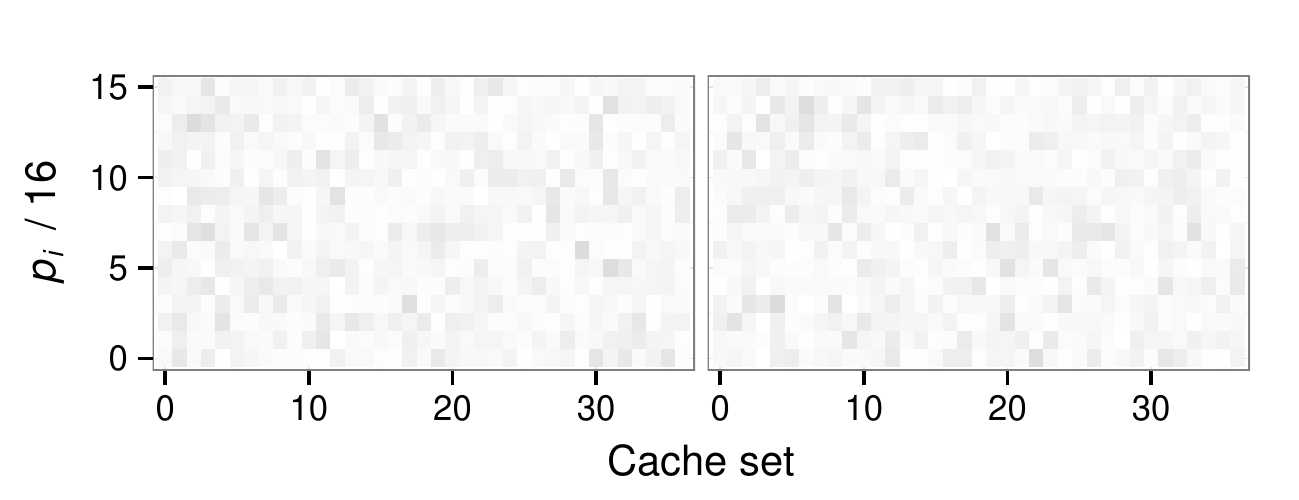}
\caption{Protecting against cache attacks on software AES}
\label{f:aes_cache_attack}
\end{figure}

We use a hardware performance monitoring counter that counts L2 cache misses 
as the probe measurement, and for each measurement we
subtract off the average measurement for that cache set for all values
of $p_i$. Figure~\ref{f:aes_cache_attack}(A) and Figure~\ref{f:aes_cache_attack}(B)
show the probe measurements when performing this attack for all values of the top 4 bits of $p_0$ (left)
and $p_5$ (right) with and without our protection scheme, respectively.
Darker cells indicate elevated measurements, and hence imply cache sets
that contain a line loaded by the attacker during the ``prime" phase that was evicted
by the AES encryption. The secret key $k$ is randomly chosen, except that $k_0 =
0$ and $k_5 = 80_{dec}$.  
Without our solution, the cache set accesses show a pattern revealing $p_i \oplus k_i$ which can be used to
determine that the top 4 bits of $k_0$ and $k_5$ are indeed $0$ and $5$,
respectively. Our solution flushes the L2 cache lazily before handing it over to any untrusted process and thus 
ensures that no signal is observed by the attacker as shown in Figure~\ref{f:aes_cache_attack}(B).

\subsection{Performance evaluation}
\label{perf_eval}
\paragraphbe{Performance costs of individual components.}
Table~\ref{tablemicrooverheads} shows the individual cost of the different 
components of our defense. Our total performance overhead is less than 
the total sum of these components as we do not perform most of these operations 
in the critical path. Note that retrieving the number of times a 
process was interrupted or determining whether a voluntary context switch 
occurred during a protected function's execution is negligible due to our
modifications to the Linux kernel described in Section~\ref{s:implementation}.

\begin{table}
\small
\centering
\begin{tabular}{p{2.5in}r}
\hline
Component & Cost (ns) \\
\hline
$m=5$ time randomization step, WCET & 710 \\
Get interrupt counters & 16 \\
Detect context switch & 4 \\
\hline
Set and restore \rtfifo & 2,650 \\
Set and restore CPU affinity & 1,235 \\
Flush L1D+L2 cache & 23,000\\
Flush BTB cache & 7,000\\
\hline
\end{tabular}
\caption{Performance overheads of individual components of our defense.
WCET indicates worst-case execution time. Only costs listed in the upper half of the table
are incurred on each call to a protected function.}
\label{tablemicrooverheads}
\end{table}

\paragraphbe{Microbenchmarks: cryptographic operations in multiple languages.}
We perform a set of microbenchmarks that test the impact of our solution
on individual operations such as RSA and ECDSA signing in the OpenSSL C
library and in the BouncyCastle Java library. In order to apply our
defense to BouncyCastle, we constructed JNI wrapper functions that call
the \fixedtimebegin and \fixedtimeend functions. Since both libraries 
implement RSA blinding to defend against timing attacks, we disable 
RSA blinding when applying our defense.

The results of the microbenchmarks are shown in Table~\ref{f:overhead-micro}.
Note that the delays experienced in any real applications will be significantly less 
than these micro benchmarks as real applications will also perform some I/O 
operations that will amortize the performance overhead.  

For OpenSSL, our solution adds between 3\% (for RSA) and 71\% (for
ECDSA) to the cost of computing a signature on average. However, we
offer significantly reduced tail latency for RSA signatures. This behavior
is caused by the fact that OpenSSL regenerates the blinding factors every
32 calls to the signing function to amortize the performance cost
of generating the blinding factors.

Focusing on the BouncyCastle results, our solution results in a 2\% decrease in
cost for RSA signing and a 63\% increase in cost for ECDSA signing,
compared to the stock BouncyCastle implementation. We believe that this increase in cost for ECDSA is 
justified by the increase in security, as the stock BouncyCastle implementation does not defend against 
local timing attacks. Furthermore, we believe that some optimizations, such
as configuring the Java VM to schedule garbage collection outside of
protected function executions, could reduce this overhead.

\begin{table}
\small
\centering
\begin{tabular}{llrc}
\hline
RSA 2048-bit sign & Mean (ms) & 99\% Tail \\
\hline
OpenSSL w/o blinding & 1.45 & 1.45 \\
Stock OpenSSL & 1.50 & 2.18 \\
OpenSSL + our solution & 1.55 & {\bf 1.59} \\
BouncyCastle w/o blinding & 9.02 & 9.41 \\
Stock BouncyCastle  & 9.80 & 10.20 \\
BouncyCastle + our solution  & {\bf 9.63} & {\bf 9.82} \\
\hline
ECDSA 256-bit sign & Mean (ms) & 99\% Tail\\
\hline
Stock OpenSSL  & 0.07 & 0.08 \\
OpenSSL + our solution & 0.12 & 0.38 \\
Stock BouncyCastle & 0.22 & 0.25 \\
BouncyCastle + our solution & 0.36 & 0.48 \\
\hline
\end{tabular}
\caption{Impact on performance of signing
a 100 byte message
using SHA-256 with RSA or ECDSA
for the OpenSSL and BouncyCastle implementations.
Measurements are in milliseconds. We disable
blinding when applying our defense to the RSA signature operation.
Bold text indicates a measurement where our defense results in better
performance than the stock implementation.
}
\label{f:overhead-micro}
\end{table}

\paragraphbe{Macrobenchmark: protecting the TLS state machine.}
\label{s:protect-tls-state}
We applied our solution to protect the server-side implementation of the TLS
connection protocol in OpenSSL. The TLS protocol is implemented as a
state machine in OpenSSL, and this presented a challenge for applying
our solution which is defined in terms of protected functions.
Additionally, reading and writing to a socket is interleaved with
cryptographic operations in the specification of the TLS protocol,
which conflicts with our
solution's requirement that no blocking I/O may be performed within a
protected function.

We addressed both challenges by generalizing the notion of a protected
function to that of a {\bf protected interval}, which is an interval of
execution starting with a call to \fixedtimebegin and ending with
\fixedtimeend. We then split an execution of the TLS protocol into
protected intervals on boundaries defined by transitions of the TLS
state machine and on low-level socket read and write operations. To
achieve this, we first inserted calls to \fixedtimebegin and \fixedtimeend
at the start and end of each state within the TLS state machine
implementation. Next, we
modified the low-level socket read and socket write OpenSSL wrapper
functions to end the current interval, communicate with the socket, and then start
a new interval. Thus divided, all cryptographic operations performed
inside the TLS implementation are within a protected interval. 
Each interval is uniquely identifiable by the name of the current TLS
state concatenated with an integer incremented every time a new interval
is started within the same TLS state (equivalently, the number of
socket operations that occurred so far during the state.)

The advantage of this strategy is that, unlike any prior defenses,  it protects the entire 
implementation of the TLS state machine from any form of timing attack. 
However, such protection schemes may incur additional overheads due to protecting parts 
of the protocol that may not be vulnerable to timing attacks because they do not
work with secret data. 

We evaluate the performance of the fully protected TLS state machine as
well as an implementation that only protects the public key signing
operation. The results are shown in Table~\ref{f:macro-overhead}. We observe an overhead of
less than 5\% on connection latency even when protecting the full TLS
protocol. 

\begin{table}
\small
\centering
\begin{tabular}{p{1.7in}llrrrr}
\hline
Connection latency (RSA) & Mean (ms) & 99\% Tail\\
\hline
Stock OpenSSL & 5.26 & 6.82 \\
Stock OpenSSL+ Our solution (sign only) & 5.33 & {\bf 6.53} \\
Stock OpenSSL+ Our solution & 5.52 & {\bf 6.74} \\
\hline
\end{tabular}
\begin{tabular}{p{1.7in}llrrrr}
\hline
Connection latency (ECDSA) & Mean (ms) & 99\% Tail\\
\hline
Stock OpenSSL & 4.53 & 6.08 \\
Stock OpenSSL+ Our solution (sign only) & 4.64 & 6.18 \\
Stock OpenSSL+ Our solution & 4.75 & 6.36 \\
\hline
\end{tabular}
\caption{
The impact on TLS v1.2 connection latency when applying our defense to
the OpenSSL server-side TLS implementation. We evaluate the cases where the the server uses an RSA 2048-bit
or ECDSA 256-bit signing key with SHA-256 as the digest function.
Latency given in milliseconds and measures the end-to-end connection
time. The client uses the unmodified OpenSSL library attempts.
We evaluate our defense when only
protecting the signing operation and when protecting all server-side
routines performed as part of the TLS connection protocol that use
cryptography.
Even when the full TLS protocol is protected, our approach adds an
overhead of less than 5\% to average connection latency.
Bold text indicates a measurement where our defense results in 
better performance than the stock implementation.
}
\label{f:macro-overhead}
\end{table}

\paragraphbe{Protecting sensitive data structures.}
We measured the overhead of applying our approach to protect the lookup
operation of the C++ STL \texttt{unordered\_map}.
For this experiment, we populate the hash map with 1 million 64-bit integer keys and values. We assume that the attacker cannot insert elements in the hash map or cause collisions. The average cost of performing a lookup of a key present in the map is
0.173$\mu$s without any defense and 2.46$\mu$s with our defense applied.
Most of this overhead is caused by the fact that the worst-case execution time of the lookup operation is significantly larger than the average-case. the profiled worst-case execution time of the lookup when interrupts do not occur is 1.32$\mu$s at $\kappa = 10^{-5}$. Thus, any timing channel defense will cause the lookup to take at least 1.32$\mu$s.  The worst-case execution estimate of the lookup operation increases to 13.3$\mu$s when interrupt cases are not excluded, hence our scheme benefits significantly from adapting to interrupts during padding for this example. Another major part of the overhead of our solution (0.710$\mu$s) comes from the randomization step to ensure safe padding . As we described earlier in Section~\ref{sec_eval}, the randomization  step is crucial to ensure that there is no timing leakage.

\paragraphbe{Hardware portability.}
Our solution is not specific to any particular hardware. It will work on
any hardware that supports standard cache hierarchy and where page coloring can be implemented.
To test the portability of our solution, we executed some of the benchmarks mentioned in Sections~\ref{sec_eval} and~\ref{perf_eval} on a 2.93 GHz Intel Xeon X5670 CPU. We confirmed that our solution successfully protects against the local and remote timing attacks on that platform too. The relative performance overheads were similar to the ones reported above.

\section{Limitations}
\paragraphbe{No system calls inside protected functions.} Our current
prototype does not support protected functions that invoke system calls.
A system call can inadvertently leak information to an attacker by leaving 
state in shared kernel data structures, which an attacker might indirectly 
observe by invoking the same system call and timing its duration. Alternatively, 
a system call might access regions of the L3 cache that can be snooped by an attacker 
process. 

The lack of system call support turned out to be not a big issue in practice as our 
experiments so far indicate that system calls are rarely used in functions dealing with 
sensitive data (e.g., cryptographic operations). However, if needed in future, one way 
of supporting system calls inside protected functions while still avoiding this leakage is 
to apply our solution to the kernel itself. For example, we can pad any system calls that 
modify some shared kernel data structures to their worst case execution times. 

\paragraphbe{Indirect timing variations in unprotected code.}
Our approach does not currently defend against timing variations in the 
execution of non-sensitive code segments that might get indirectly affected by a 
protected function's execution. For example, consider the case where a non-sensitive function from a process gets scheduled on a processor core immediately after another process from the same user finishes executing a protected function. In such a case, our solution will not flush the state of per-core resources like L1 cache as both these processes belong to the same user. However, if such remnant cache state affects the timing of the non-sensitive function, an attacker may be able to observe these variations and infer some information about the protected function.

Note that currently there are no known attacks that could exploit this kind of leakage.
A conservative approach that prevents such leakages is to flush 
all per-cpu resources at the end of each protected function. This will, of course, result in higher performance overheads. The costs associated with cleansing different types of per-cpu resources are summarized in Table~\ref{tablemicrooverheads}.

\paragraphbe{Leakage due to fault injection.}
If an attacker can cause a process to crash in the middle of a protected function's execution, the attacker can potentially learn secret information. For example, consider a protected function that first performs a sensitive operation and then parses some input from the user. An attacker can learn the duration of the sensitive operation by providing a bad input to the parser that makes it crash and measuring how long it takes the victim process to crash.

Our solution, in its current form, does not protect against such attacks. However, this is not a fundamental limitation. One simple way of overcoming these attacks is to modify the OS to apply the time padding for a protected function even after it has crashed as part of the OS's cleanup handler. This can be implemented by modifying the OS to keep track of all processes that are executing protected functions at any given point of time and their  respective padding parameters. If any protected function crashes, the OS cleanup handler for the corresponding process can apply the desired amount of padding.

\section{Related work}  \label{sec:related}
\subsection{Defenses against remote timing attacks}
The remote timing attacks exploit the input-dependent execution 
times of cryptographic operations. There are three main approaches to 
make cryptographic operations' execution times independent of their inputs:
static transformation, application-specific changes, and dynamic padding. 

\paragraphbe{Application-specific changes.}
One conceptually simple way to defend an application
against timing attacks is to
modify its sensitive operations such that their timing behavior is not
key-dependent. For example, AES~\cite{konighofer2008fast, blomer2005provably, kasper2009faster}
implementations can be  modified to ensure that
their execution times are key-independent. Note that, since the cache
behavior impacts running time, achieving secret-independent timing
usually requires rewriting the operation so that its memory access
pattern is also independent of secrets. Such modifications are application specific, hard to design, 
and very brittle. By contrast, our solution is completely independent of the application and the programming 
language. 

\paragraphbe{Static transformation.} An alternative approach to prevent remote attacks is 
to use static transformations on the implementation of the cryptographic operation
to make it constant time. One can use a static analyzer to
find the longest possible path through the cryptographic operation and insert padding 
instructions that have no side-effects (like NOP) along other paths so that they take 
the same amount of time as the longest path~\cite{van2012compiler, coppens2009practical}.
While this approach is generic and can be applied to any sensitive operation, it has 
several drawbacks. In modern architectures like x86, the execution time of several instructions 
(e.g., the integer divide instruction and multiple floating-point instructions) depend the value 
of the input of these instructions. This makes it extremely hard and time consuming to 
statically estimate the execution time of these instructions. Moreover, it is very hard to statically predict 
the changes in the execution time due to internal cache collisions in the implementation of the 
cryptographic operation. To avoid such cases, in our solution, we use dynamic offline profiling 
to estimate the worst-case runtime of a protected function. However, such dynamic techniques 
suffer from incompleteness i.e. they might miss worst-case execution times triggered by pathological 
inputs. 

\paragraphbe{Dynamic padding.} Dynamic padding techniques add a variable
amount of padding to a sensitive computation that depends on the
observed execution time of the computation in order to mitigate the
timing side-channel. Several
prior works~\cite{askarov2010predictive, zhang2011predictive,
Cock_GMH_14, kopf2009provably, fuzzdb} have presented ways to pad the execution of a black-box 
computation to certain predetermined thresholds and obtain bounded
information leakage. 
Zhang et al. designed a new programming language that, when used to write 
sensitive operations, can enforce limits on the timing information leakage~\cite{zhang2012language}.
The major drawback of existing dynamic padding schemes is that they incur large performance 
overhead. This results from the fact that their estimations of the worst-case execution time tend to be
overly pessimistic as it depends on several external parameters like OS scheduling, cache behavior of the 
simultaneously running programs, etc. For example, Zhang et al.~\cite{zhang2011predictive} set the 
worst-case execution time to be $300$ seconds for protecting a Wiki server. Such overly pessimistic 
estimates increase the amount of required padding and thus results in significant performance overheads
($90-400\%$ in macro-benchmarks~\cite{zhang2011predictive}). Unlike existing dynamic padding schemes, our solution 
incurs minimal performance overhead and protects against both local and remote timing attacks.

\subsection{Defenses against local attacks}
Local attackers can also perform timing attacks, hence some of the defenses provided in the prior section 
may also be used to defend against some local
attacks. However, local attackers also have access to shared hardware resources that contain 
information related to the target sensitive operation. The local attackers also have access 
to fine-grained timers.

A common local attack vector is to probe a shared hardware resource, and then,
using the fine-grained timer, measure how long the probe took to run.
Most of the proposed defenses to such attacks try to either remove access to
fine-grained timers or isolate access to the shared hardware resources. Some 
of these defenses also try to minimize information leakage by obfuscating the 
sensitive operation's hardware access patterns. We describe 
these approaches in detail below.

\paragraphbe{Removing fine-grained timers.}
Several prior projects have evaluated removing or modifying time
measurements taken on the target machine~\cite{martin2012timewarp, li2013mitigating,vattikonda2011eliminating}.
Such solutions are often quite effective at preventing a large number of local side channel attacks as the underlying states of
most shared hardware resources can only be read by accurately measuring the time taken to perform
certain operations (e.g., read a cache line). 

However, removing access to wall clock time is not sufficient for protecting 
against all local attackers. For example, a local attacker executing multiple probe threads can infer time measurements by
observing the scheduling behavior of the threads. Custom scheduling schemes (e.g., instruction-based scheduling) can eliminate such an attack~\cite{stefan2013eliminating} but implementing these defenses require major changes to the OS scheduler. In contrast, our solution only requires minor changes to the OS scheduler and protects against both local and remote attackers.  

\paragraphbe{Preventing sharing of hardware state across processes.}
Many proposed defenses against local attackers  prevent an attacker from 
observing state changes to shared hardware resources caused by a victim 
process. We divide the proposed defenses into five categories and describe them next.

\paragraphbe{Resource partitioning.} Partitioning shared hardware
resources can defeat local attackers, as they cannot access the same
partition of the resource as a victim.
Kim et al.~\cite{kim2012system} present an efficient management
scheme for preventing local timing attacks across virtual machines (VMs). Their 
technique locks memory regions accessed by sensitive functions 
into reserved portions of the L3 cache. This scheme can be more 
efficient than page coloring. Such protection schemes are complementary 
to our technique. For example, our solution can be modified to use such a 
mechanism instead of page coloring to dynamically partition the L3 cache. 

Some of the other resource partitioning schemes (e.g., Ristenpart et al.~\cite{ristenpart2009hey}) suggest allocating dedicated hardware to each 
virtual machine instance to prevent cross-VM attacks. However, 
such schemes are wasteful of hardware resources as they decrease the amount of resources available to concurrent processes. By contrast, our solution utilizes the shared 
hardware resources efficiently as they are only isolated during the execution of 
the protected functions. The time a process spends executing protected functions 
is usually much smaller than the time it spends in non-sensitive computations. 

\paragraphbe{Limiting concurrent access.} If gang scheduling~\cite{kim2012system} is used or hyperthreading is disabled, an attacker can only observe per-CPU resources 
when it has preempted a victim. Hence, reducing the frequency of preemptions reduces the feasibility of cache-attacks on per-CPU caches. Varadarajan et al.~\cite{varadarajan2014scheduler} propose using minimum runtime
guarantees to ensure that a VM is not preempted too frequently. However,
as noted in~\cite{varadarajan2014scheduler}, such a scheme is very hard
to implement in a OS scheduler as, unlike a hypervisor scheduler, an OS
scheduler must deal with a unbounded number of processes.

\paragraphbe{Custom hardware.} Custom hardware can be used to obfuscate and 
randomize the victim process's usage of the hardware. For example,  
Wang et al.~\cite{wang2007new, wang2008novel} proposed new 
ways of designing caches that ensures that no information about cache 
usage is shared across different processes. However such schemes have 
limited practical usage as they, by design, cannot be deployed on off-the-shelf 
commodity hardware.

\paragraphbe{Flushing state.} Another class of defenses ensure that the
state of any per-CPU hardware resources are cleared before transferring
them from one process to another. D\"uppel, by Zhang et al.~\cite{zhang13}, 
flushes per-CPU L1 and (optionally) L2 caches periodically in a multi-tenant VM 
setting. Their solution also requires the hyperthreading to be disabled. 
They report around $7$\% overheads on regular workloads. In essence, this scheme
is similar to our solution's technique of flushing per-CPU resources in the 
OS scheduler. However, unlike D\"uppel, we flush the state lazily only when a 
context switch to a different user process than the one executing a protected 
operation occurs.  Also, D\"uppel only protects against local cache attacks. We 
protect against both local and remote timing and cache attacks while still incurring 
less overhead than D\"uppel. 

\paragraphbe{Application transformations.} Sensitive operations like sensitive computations in different programs can also be modified to exhibit either secret-independent or obfuscated hardware access patterns. If the access to the hardware is independent of secrets, then an attacker cannot use any of the state leaked through shared hardware to learn anything meaningful about the sensitive operations. Several prior projects have shown how to modify AES implementations to obfuscate their cache access patterns~\cite{bernstein2005cache, 
blomer2005provably,brickell2006software, tromer2010efficient, osvik2006cache}. Similarly, recent versions of OpenSSL use a specifically modified implementation of RSA  that ensures secret-independent cache accesses. Some of these transformations can also be applied dynamically. For example, Crane et al.~\cite{crane2015thwarting} implement a system that dynamically applies cache-access obfuscating transformations to an application at runtime. 

However, these transformations are specific to particular cryptographic operations and are very hard to implement and maintain correctly. For example, $924$ lines of assembly code had to be added to OpenSSL to implement make the RSA implementation's cache accesses secret-independent.

\section{Conclusion}

We presented a low-overhead, cross-architecture defense that protects applications 
against both local and remote timing attacks with minimal application code changes.  
Our experiments and evaluation also show that our defense works across different 
applications written in different programming languages. 

Our solution defends against both local and remote attacks by using a combination 
of two main techniques: (i) a time padding scheme that only takes secret-dependent 
time variations into account, and (ii) preventing information leakage via shared resources
such as the cache and branch prediction buffers.  We demonstrated that applying small
time pads accurately is non-trivial because the timing loop itself may leak information.  
We developed a method by which small time pads can be applied securely. We hope that 
our work will motivate application developers to leverage some of our techniques to protect 
their applications from a wide variety of timing attacks. We also expect that the underlying 
principles of our solution will be useful in future work protecting against other forms of 
side channel attacks.

\section*{Acknowledgments}

This work was supported by NSF, DARPA, ONR, and a Google PhD
Fellowship to Suman Jana.  Opinions, findings and conclusions or
recommendations expressed in this material are those of the author(s)
and do not necessarily reflect the views of DARPA.

\setlength{\bibsep}{2.5pt}
\small
\bibliographystyle{abbrv}
\bibliography{fixedTime}

\end{document}